# Reduced decoherence using squeezing, amplification, and anti-squeezing

R.A. Brewster, T.B. Pittman, and J.D. Franson
University of Maryland Baltimore County, Baltimore, MD 21250 USA

**Abstract**: Loss and decoherence are a major problem in the transmission of non-classical states of light over large distances. It was recently shown that the effects of decoherence can be reduced by applying a probabilistic noiseless attenuator before transmitting a quantum state through a lossy channel, followed by probabilistic noiseless amplification (M. Micuda et al, Phys. Rev. Lett. **109**, 180503 (2012)). Here we show that similar results can be obtained for certain kinds of macroscopic quantum states by squeezing the signal before transmission, followed by deterministic amplification and anti-squeezing to restore the original amplitude of the state. This approach can greatly reduce the effects of decoherence in the transmission of non-Gaussian states, such as Schrodinger cat states, without any reduction in the data transmission rate.

## I. INTRODUCTION

Schrodinger cats [1-2] and other macroscopic superposition states are very susceptible to the effects of loss [3-5], which makes it difficult to transmit them over large distances. Here we show that the decoherence of Schrodinger cat states can be greatly reduced by applying an appropriate squeezing operation [6-19] before their transmission through a lossy medium [20-21], followed by deterministic amplification and anti-squeezing [18-19, 22] to restore the original amplitude of the state. This process can reduce the amount of decoherence in non-Gaussian macroscopic states by many orders of magnitude while maintaining the original data rate.

Micuda et al. [23] previously proposed a somewhat similar technique in which the amplitude of the signal is attenuated using a probabilistic noiseless attenuator before transmission, followed by a probabilistic noiseless amplifier to restore the original amplitude as shown in Fig. 1(a). Noiseless attenuation [23-26] and noiseless amplification [27-29] can both be implemented using various post-selection and heralding techniques. The output of the system is only accepted when certain conditions are met, which reduces the data transmission rate exponentially. Thus the decreased decoherence is achieved at the cost of a reduced data rate.

The exponential decrease in the data rate can be avoided for Schrodinger cat states by applying squeezing [6-19], deterministic amplification, and anti-squeezing [18-19, 22] operations instead, as illustrated in Fig. 1(b). An incident Schrodinger cat state is first squeezed in such a way as to reduce the overall amplitude of its two phase components as illustrated in Fig. 2. After passing through a lossy channel, the signal is amplified using a deterministic amplifier, such as an optical parametric amplifier (OPA) [30-32], and then restored to its original amplitude by applying an appropriate anti-squeezing operation. The anti-squeezing operation $\hat{S}^\dagger(r)$ is the inverse of the squeezing operation $\hat{S}(r)$, where $r$ is the usual squeezing parameter as defined below. We will show that this process can reduce the decoherence by many orders of magnitude under the appropriate conditions. Similar results are expected for other kinds of macroscopic superposition states.

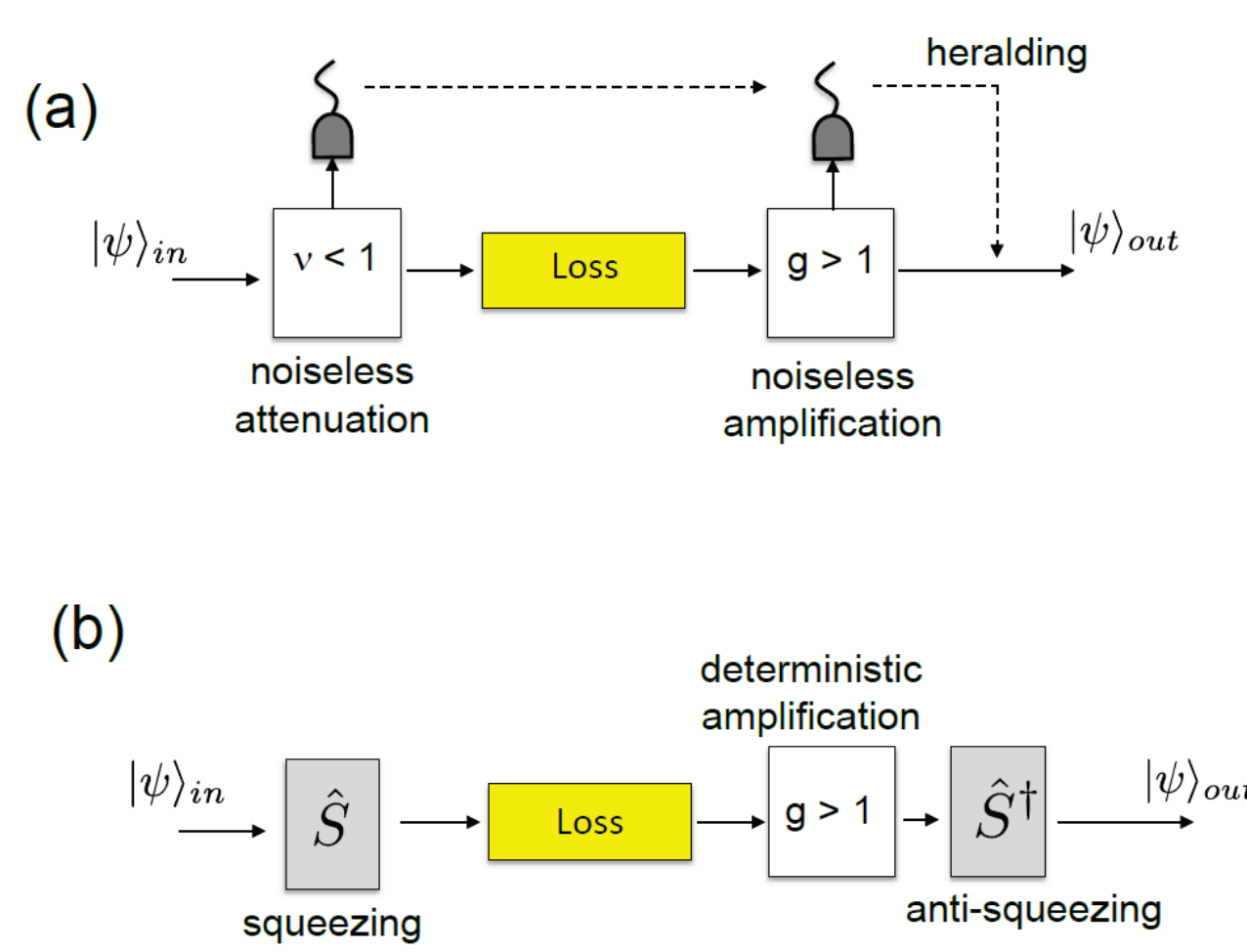

**Fig. 1.** (a) Reduction in the decoherence of a quantum signal by applying a noiseless attenuation factor $\nu$ before transmission through a lossy channel, followed by noiseless amplification with gain $g$ [23]. The probabilistic nature of the noiseless attenuation and amplification results in an exponential decrease in the data transmission rate in this approach. (b) Reduction in the decoherence of a Schrodinger cat state by applying an appropriate squeezing operation $\hat{S}$ before transmission, followed by a deterministic amplifier (OPA) and anti-squeezing $\hat{S}^\dagger$. This approach has the advantage that all of the operations are deterministic and the data transmission rate is not reduced as a result.

A recent experiment by Le Jeannic et al. showed that squeezing a Schrodinger cat state can help to maintain the negative part of its Wigner distribution in the presence of loss [21], as was first suggested by R. Filip [20]. These earlier papers did not include the effects of amplification, however, which is required to restore the original amplitude of the cat state.

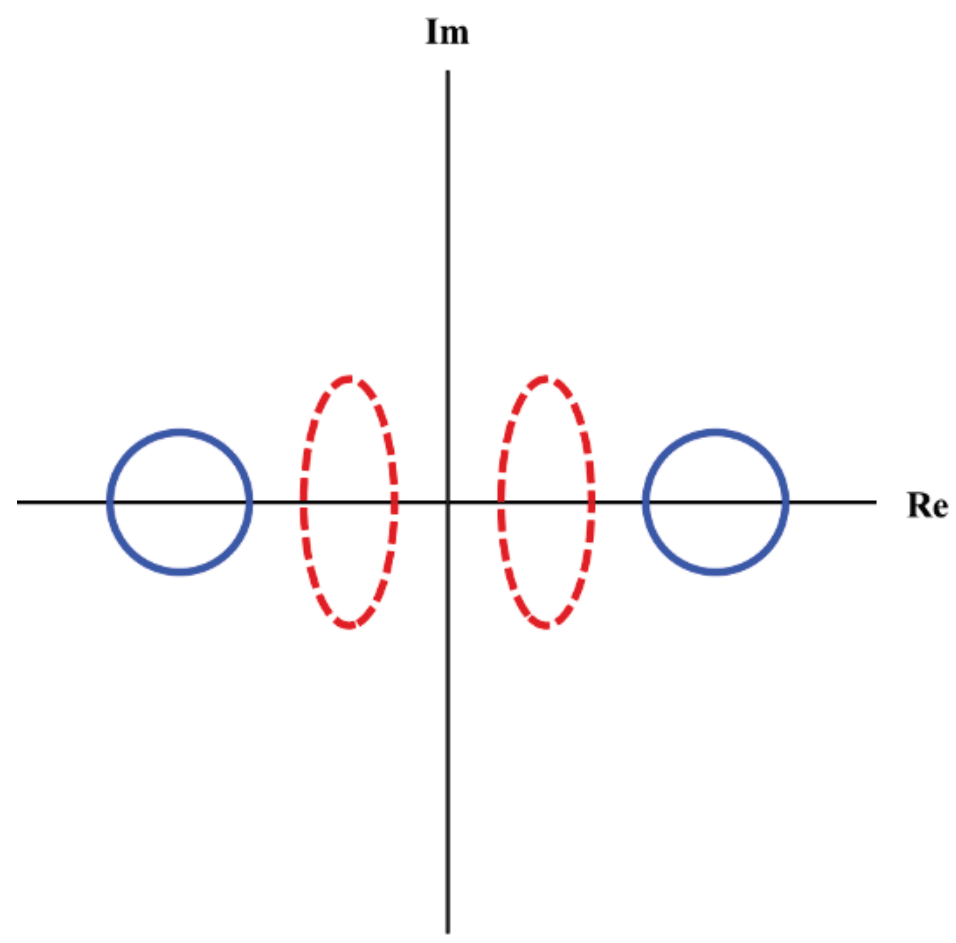

**Fig. 2:** (Color online) Phase-space diagram illustrating the squeezing of a Schrodinger cat state before transmission through a lossy channel to minimize decoherence as shown in Fig. 1. The real and imaginary axes labelled Re and Im correspond to either the Wigner distribution or the Q-function, while the solid and dashed lines represent the 1- $\sigma$ contours of the relevant Gaussian distributions. The initial cat state is assumed to have components that differ by a phase shift of $\pi$ as indicated by the blue (solid) lines. The squeezing parameters are chosen to reduce the amplitudes along the real axis as indicated by the red (dashed) lines. This process decreases the overall amplitude of the signal and thus the number of photons left in the environment due to loss and amplification.

One of the goals of this paper is to include the effects of an amplifier in addition to loss. It will be found that the residual decoherence due to amplification is comparable to that due to loss when the cat state is restored to its original amplitude. We provide an analytic solution to this problem using the Q-function quasiprobability distribution. This provides insight into the optimal amount of squeezing, which will be found to minimize the number of idler photons generated in the OPA. Another goal is to investigate trade-offs between the relevant physical parameters, which is necessary to minimize the overall decoherence and determine the expected performance of the system.

Niset et al. have proven a no-go theorem which shows that Gaussian operations, such as squeezing, cannot protect Gaussian states from decoherence [33]. The Wigner distribution of a Schrodinger cat state is not a Gaussian and this no-go theorem does not apply to our approach. The use of squeezing and anti-squeezing in this way is limited to non-Gaussian states, however, and there are limits on the amount of decoherence reduction that can be achieved.

Section II begins by discussing a Schrodinger cat interferometer that can be used to measure the amount of quantum coherence between the two components of the cat state after loss and amplification. An analytic solution for the visibility of the quantum interference is calculated in Section III using the Husimi-Kano Q-function [34-35]. The results of the calculations are described for a range of parameters in Section IV, which includes a comparison of the effects of loss versus amplification. A summary and conclusions are provided in Section V.

## II. SCHRODINGER CAT INTERFEROMETER

The reduction in the decoherence can be observed using the Schrodinger-cat interferometer of Fig. 3 [3-5]. A Schrodinger cat state is probabilistically generated using the source enclosed in the dashed lines on the left, starting from a coherent state $|\alpha_0\rangle$ with complex amplitude $\alpha_0$ as described in the figure caption. After passing through the squeezer, transmission channel, amplifier, and anti-squeezing operations, the coherence of the resulting state can be measured by looking for quantum interference between the two components of the original cat state using the analyzer enclosed in the dashed lines on the right-hand side of Fig. 3. The Schrodinger cat interferometer itself was described in more detail in Refs. [2-5].

A Schrodinger cat state is created [1-2] in the source box on the left by passing an initial coherent state $|\alpha_0\rangle$ through a Kerr medium $K$ that is located in one path of a single-photon interferometer. The Kerr medium is assumed to produce a phase shift of $2\phi$ if single photon $\gamma_1$ passes through it. By applying a constant phase shift of $-\phi$ and post-selecting on the detection of $\gamma_1$ in the detector shown, this process will produce a cat state whose components have been shifted by $\pm\phi$ depending on the path taken by $\gamma_1$, as illustrated in Fig. 2 for $\phi = \pi / 2$.

After passing through a squeezer, lossy transmission channel, OPA, and an anti-squeezer, the visibility of the quantum interference between the two components of the cat state can be measured using the apparatus shown in the box on the right of Fig. 3. Here a second phase shift of $\pm\phi$ is applied depending on the path taken by photon $\gamma_2$, with post-selection based on the detection of $\gamma_2$ in the detector shown. The phase of the signal is then measured with a homodyne detector and the events are further post-selected based on a measured phase shift of approximately zero. With a net phase shift of zero, the two components of the original cat state will now overlap in phase space as illustrated in more detail in Fig. 4. Quantum interference between these two probability amplitudes will then occur with a visibility that that depends on their degree of coherence.

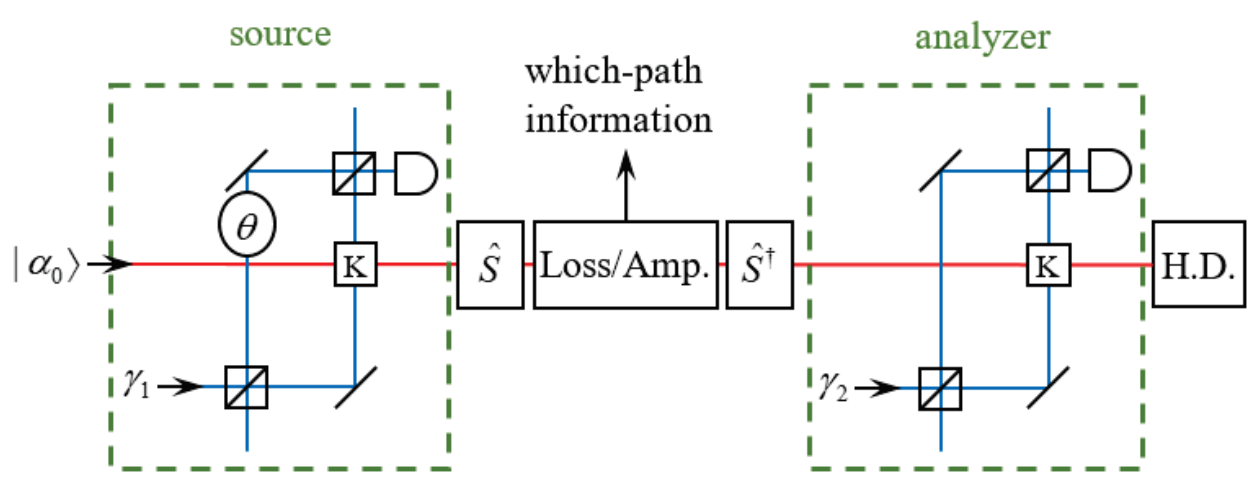

**Fig. 3**: (Color online) Measurement of the amount of decoherence using a Schrodinger cat interferometer [3-5]. The source box on the left produces a Schrodinger cat state as illustrated in Fig. 2. Here $|\alpha_0\rangle$ is an initial coherent state and $K$ is a Kerr medium located in one path of a single-photon interferometer. The Kerr medium produces a phase shift of $2\phi$ if single photon $\gamma_1$ passes through it. A constant phase shift of $-\phi$ is applied in both paths (not shown), which gives a net phase shift of $\pm\phi$ depending on the path taken by $\gamma_1$. A variable phase shift $\theta$ is also applied to $\gamma_1$ in one path of the single-photon interferometer, with post-selection on the detection of $\gamma_1$ in the detector shown. After passing through a squeezer $\hat{S}$, lossy transmission channel, OPA, and an anti-squeezer $\hat{S}^\dagger$, the visibility of the quantum interference between the two components of the cat state is measured using the apparatus shown in the analyzer box on the right. Here a second photon $\gamma_2$ passes through a single-photon interferometer with a Kerr medium in one of its paths, which produces another phase shift of $\pm\phi$. The events are post-selected based on single-photon detection in the detectors shown, along with a net phase shift of zero as measured by the homodyne detector (H.D.). This results in quantum interference between the two components of the original cat state, as described in more detail in Fig. 4.

In the limit of $|\alpha_0| >> 1$, most of the decoherence during transmission is due to which-path information left in the environment. For example, passing a cat state through a beam splitter (a common model for loss) will produce a second coherent state in the other output port of the beam splitter, with a phase that is different for the two components of the cat state [1-5]. As a result, entanglement between the components of the cat state and the beam splitter output will substantially reduce the quantum interference. Entanglement between the signal and idler modes of an OPA will also produce which-path information of this kind, which can be the dominant source of decoherence in a linear amplifier [5]. In either case, the squeezing operation of Fig. 2 reduces the overall amplitude of the cat state components during transmission and amplification, which reduces the number of photons left in the environment and thus the which-path information.

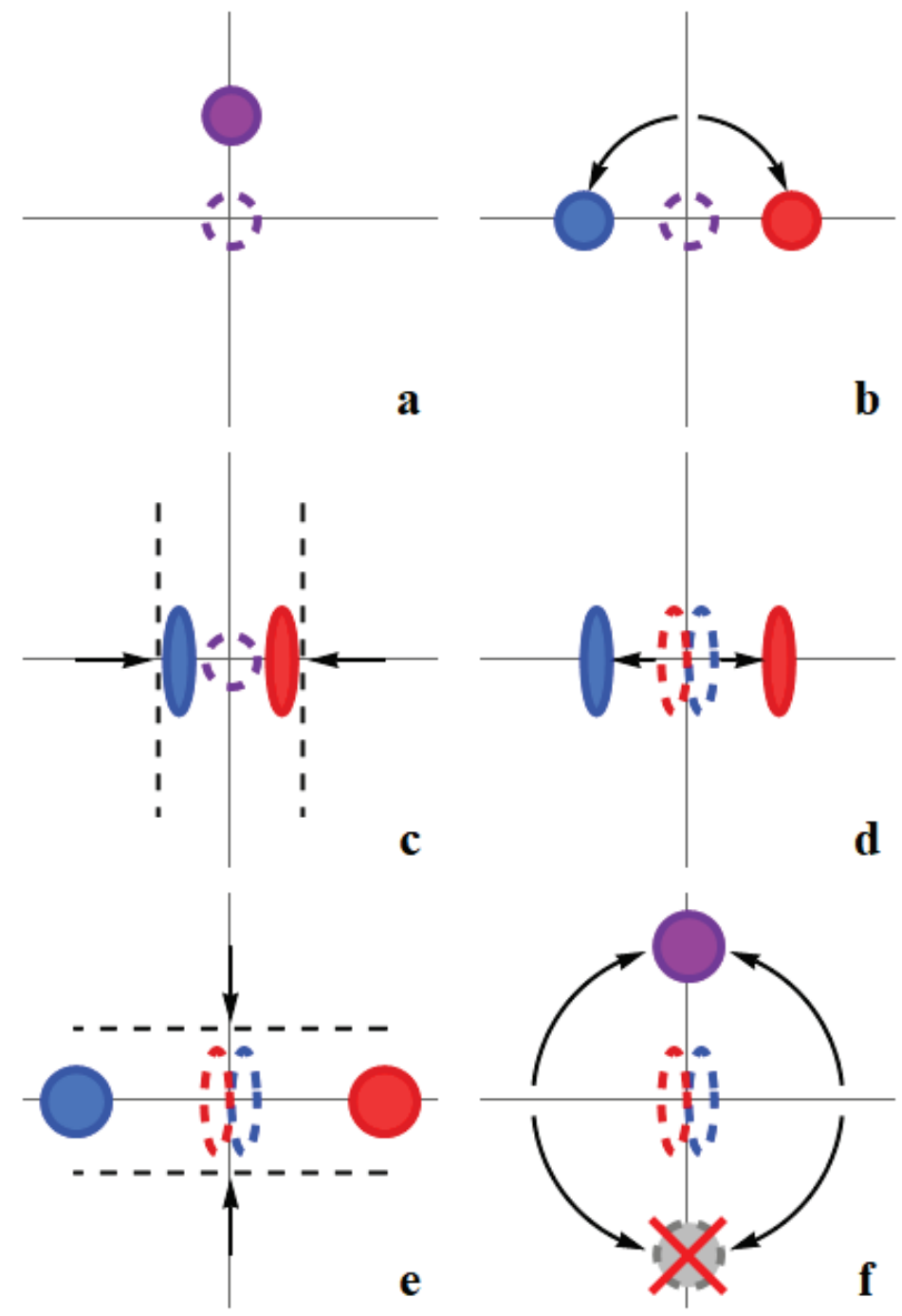

**Fig. 4**: (Color online) Phase-space representation of the state of the system as it progresses through the apparatus shown in Fig. 3 for the case of $\phi = \pi/2$. The horizontal axes correspond to the real part of the Wigner distribution or Q-function as in Fig. 2, while the vertical axes correspond to the imaginary part. (a) The initial coherent state $|\alpha_0\rangle$ (solid circle) along with the idler mode of the OPA (dotted circle) which is initially in its vacuum state. (b) Cat state created by the first single-photon interferometer, where the Kerr medium and a constant phase shift apply a net phase shift of $\pm\pi/2$ depending on the path taken by the single photon. (c) Reduced overall amplitude of the cat state components due to squeezing along the real axis. The squeezing also increases the amplitude along the imaginary axis. (d) Compensation for the effects of loss using a deterministic amplifier, which also displaces the idler modes. This results in a state where the signal and idler modes are entangled. (e) Restoration of the original amplitude of the cat state using anti-squeezing. (f) A second phase shift of $\pm\pi/2$ is produced in the analyzer of Fig. 3, depending on the path taken by the second single photon. Post-selecting on a net phase shift of zero results in an overlap between the two original components of the cat state. This gives quantum interference, whose visibility provides a measure of the amount of decoherence, as described in more detail in Ref. [5].

## III. ANALYSIS USING THE Q-FUNCTION

The visibility of the quantum interference can be calculated analytically using the Husimi-Kano Q-function [34-35]. After the first post-selection process, the initial state $|\psi\rangle_{in,s}$ of the Schrodinger cat is given by

$$|\psi\rangle_{\rm in,s}=\frac{1}{2}\left(|\alpha_0 e^{i\phi}\rangle+e^{i\theta}\,|\alpha_0 e^{-i\phi}\rangle\right). \quad (1)$$

The factor of 1/2 comes from the post-selection and the $e^{i\theta}$ results from the phase shift $\theta$ inserted into one arm of the first single-photon interferometer in Fig. 3. The state in Eq. (1) is not normalized and its norm reflects the probability of achieving that output, as will the norms of subsequent post-selected states. The factor of 1/2 does not affect the calculated visibility but it is useful in calculating the probability of success for the post-selection process.

The four terms in the corresponding density operator can be written as $\hat{\rho}=\hat{\rho}_{++}+\hat{\rho}_{+-}+\hat{\rho}_{-+}+\hat{\rho}_{--}$, where

$$\hat{\rho}_{+-}=\left|e^{i\phi}\alpha_0\right\rangle\left|0_i\right\rangle\left\langle 0_i\right|\left\langle e^{-i\phi}\alpha_0\right|/4 \quad (2)$$

for example, with a similar notation for the other three terms [5]. Here the $\pm$ signs correspond to the sign of the phase shift $\pm\phi$ in the original cat state of Eq. (1). We have assumed that the idler mode of the OPA is initially in its vacuum state $\left|0_i\right\rangle$.

The single-mode squeezing operation produces a unitary transformation $\hat{S}(r)$ given by [18-19]

$$\hat{S}(r)\equiv e^{r(e^{-i\xi}\hat{a}^2-e^{i\xi}\hat{a}^{\dagger 2})/2}, \quad (3)$$

where $\hat{a}$ is the photon annihilation operator for the signal field. Here $r$ is the usual squeezing parameter which depends on the coupling between the pump and signal, the interaction time, and the phase $\xi$ of the pump [19]. Eq. (3) can be factored into a more useful form given by [37]

$$\hat{S}=\frac{1}{\sqrt{\mu}}e^{-\sqrt{\mu^2-1}e^{i\xi}\hat{a}^{\dagger 2}/2\mu}\mu^{-\hat{a}^\dagger\hat{a}}e^{\sqrt{\mu^2-1}e^{-i\xi}\hat{a}^2/2\mu}, \quad (4)$$

where $\mu=\cosh r$ and we have dropped the explicit dependence of $\hat{S}(r)$ on $r$.

For simplicity, the analysis presented in the text will only include amplification using an OPA, since the effects of the amplifier is one of the main topics of interest. The more general case of loss followed by amplification gives similar results as shown in the appendices. The amplification process corresponds to a unitary transformation $\hat{U}$ given by [32, 37-38]

$$\hat{U}=\frac{1}{g}e^{-\sqrt{g^2-1}\hat{a}^\dagger\hat{b}^\dagger/g}g^{-(\hat{a}^\dagger\hat{a}+\hat{b}^\dagger\hat{b})}e^{\sqrt{g^2-1}\hat{a}\hat{b}/g}. \quad (5)$$

Here $\hat{b}$ is the annihilation operator for the amplifier's idler mode and $g=\cosh(\kappa t)$ is the gain of the amplifier, where $\kappa$ is the coupling between the pump and the signal and idler modes and $t$ is the interaction time [32].

The final single-photon interferometer in Fig. 3 performs a probabilistic phase shift that can be represented by the operator $\hat{T}$ given by [5]

$$\hat{T}=\frac{1}{2}\left(e^{i\phi\hat{a}^\dagger\hat{a}}+e^{-i\phi\hat{a}^\dagger\hat{a}}\right). \quad (6)$$

Combining Eqs. (1) through (6) allows the final state of the system in Fig. 3 to be written as

$$|\psi\rangle_{\rm out}=\hat{T}\hat{S}^\dagger\hat{U}\hat{S}\,|\psi\rangle_{{\rm in},s}\,|0\rangle_i. \quad (7)$$

This corresponds to a pure state that is a superposition of four terms, since we have not yet traced over the idler modes. Post-selection on a net phase shift of 0 in the homodyne measurement will reduce this to a superposition of two terms, since the other two terms correspond to phase shift of $\pi$ and they are eliminated as indicated by the red cross in Fig. 4f. The two remaining terms produce quantum interference between the two components of the original cat state.

The two-mode Q-function is defined by

$$Q(\alpha,\beta)\equiv\frac{1}{\pi^2}\langle\beta|\langle\alpha|\hat{\rho}|\alpha\rangle|\beta\rangle. \quad (8)$$

Here $\alpha$ and $\beta$ are arbitrary complex variables corresponding to the amplitudes of coherent states $|\alpha\rangle$ and $|\beta\rangle$ in the signal and the idler mode of the OPA, respectively, while $\hat{\rho}$ is the density operator of Eq. (2). Since the Q-function is linear in $\hat{\rho}$, the four terms in the initial density operator allow the Q-function to be written in the analogous form

$$\begin{aligned}Q(\alpha,\beta)&=Q_{++}(\alpha,\beta)+Q_{+-}(\alpha,\beta)\\&+Q_{-+}(\alpha,\beta)+Q_{--}(\alpha,\beta).\end{aligned} \quad (9)$$

After post-selection, inserting the final state of Eq. (7) into Eq. (8) gives

$$\begin{aligned}&Q_{+-}(\alpha,\beta)=\\&\frac{e^{-i\theta}}{16\pi^2}\langle\beta|_i\,\langle\alpha|_s\,e^{-i\phi\hat{a}^\dagger\hat{a}}\hat{S}^\dagger\hat{U}\hat{S}\,|\alpha_0 e^{+i\phi}\rangle_s\,|0\rangle_i\\&\times\langle 0|_i\,\langle\alpha_0 e^{-i\phi}|_s\,\hat{S}^\dagger\hat{U}^\dagger\hat{S}e^{-i\phi\hat{a}^\dagger\hat{a}}\,|\alpha\rangle_s\,|\beta\rangle_i,\end{aligned} \quad (10)$$

with similar results for the other three terms. It should be noted that both terms in the operator $\hat{T}$ in Eq. (6) are retained, but they appear separately in the four $Q_{\pm\pm}$ terms.

It will be convenient to define $f_+$ as the first factor on the right-hand side of Eq. (10):

$$f_+ \equiv \langle\beta|_i \langle\alpha|_s e^{-i\phi\hat{a}^\dagger\hat{a}}\hat{S}^\dagger(r)\hat{U}\hat{S}(r)|\alpha_0 e^{+i\phi}\rangle_s|0\rangle_i, \qquad (11)$$

with an analogous definition for $f_-$. In that case

$$Q_{+-}(\alpha,\beta) = \frac{e^{-i\theta}}{16\pi^2} f_+ f_-^*. \qquad (12)$$

The completeness property of the coherent states allows $f_+$ to be rewritten as

$$\begin{aligned} f_+ &= \frac{1}{\pi^4}\iiiint \langle\beta|_i \langle\alpha e^{+i\phi}|_s \hat{S}^\dagger|\gamma\rangle_s|\delta\rangle_i \\ &\times\langle\delta|_i\langle\gamma|_s \hat{U}|\varepsilon\rangle_s|\zeta\rangle_i \qquad (13) \\ &\times\langle\zeta|_i\langle\varepsilon|_s \hat{S}|\alpha_0 e^{+i\phi}\rangle_s|0\rangle_i d^2\gamma d^2\delta d^2\varepsilon d^2\zeta, \end{aligned}$$

where we let the operator $e^{-i\phi\hat{a}^\dagger\hat{a}}$ act to the left.

Inserting the factored forms of the operators $\hat{U}$ and $\hat{S}$ into Eq. (13) gives

$$\begin{aligned} f_+ &= \frac{e^{-(|\alpha|^2+|\beta|^2+|\alpha_0|^2)/2}}{\mu g \pi^4} e^{\sqrt{\mu^2-1}(\alpha^{*2}e^{i(\xi-2\phi)}+\alpha_0^2 e^{-i(\xi-2\phi)})/2\mu} \\ &\times\iiiint e^{-(|\gamma|^2+|\delta|^2+|\varepsilon|^2+|\zeta|^2)} e^{-\sqrt{\mu^2-1}(\gamma^2 e^{-i\xi}+\varepsilon^{*2}e^{i\xi})/2\mu} \qquad (14) \\ &\times e^{\sqrt{g^2-1}(\varepsilon\zeta-\gamma^*\delta^*)/g} e^{(\gamma^*\varepsilon+\delta^*\zeta)/g} \\ &\times e^{(\alpha^*\gamma e^{-i\phi}+\varepsilon^*\alpha_0 e^{i\phi})/g} e^{\beta^*\gamma} d^2\gamma d^2\delta d^2\varepsilon d^2\zeta. \end{aligned}$$

with an analogous expression for $f_-$.

The integrals in Eq. (14) can be evaluated analytically to calculate the Q-function and the visibility, as described in more detail in the appendices [37].

## IV. RESULTS AND COMPARISONS

Squeezing and anti-squeezing reduce the amount of decoherence by reducing the number of photons left in the environment. This effect can be seen in Fig. 5, where the visibility of the interference pattern is plotted as a function of the squeezing parameter $r$ for an initial amplitude of $|\alpha_0| = 100$. The solid red curve shows the visibility for a relatively small gain of 1.001, while the dotted blue curve corresponds to loss modeled by a beam splitter with a transmission coefficient of $t = 0.999$. (The loss and gain both refer to the change in the amplitude of the signal rather than the intensity.) The black dotted curve shows the combined effects of loss and gain. It can be seen that loss and gain have essentially the same effect on the visibility of the quantum interference under these conditions of low gain, if the gain is chosen to be $g = 1/t$ in order to restore the original signal amplitude.

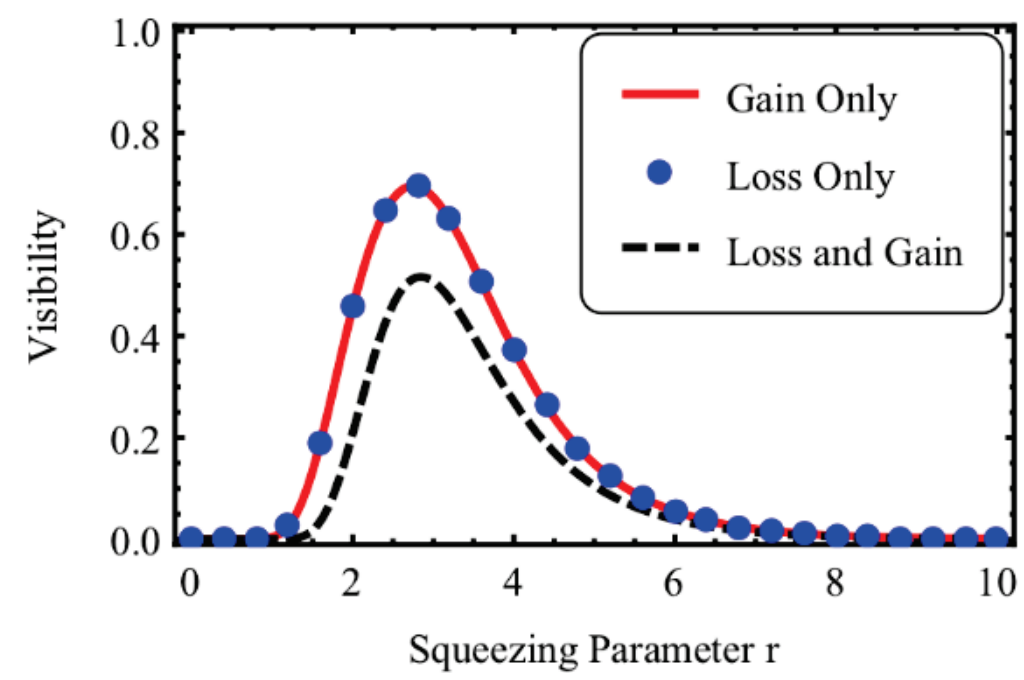

**Fig 5.** (Color online) Plot of the visibility of the quantum interference as a function of the squeezing parameter r. The solid red curve corresponds to a relatively small gain of g = 1.001 while the dotted blue line corresponds to a loss factor of $t = 0.999$. The black dashed curve shows the combined effects of both loss and gain. The amplitude of the coherent state was $|\alpha_0| = 100$ and the parameters in the interferometer of Fig. 3 were chosen to be $\phi = \pi/2$ and $\theta = 0$. These results show that squeezing and anti-squeezing can produce a large improvement in the visibility even when the loss and gain are relatively small.

Even the relatively small loss and gain shown in Fig. 5 will produce an exponentially large reduction in the visibility for $|\alpha_0| = 100$ in the absence of any squeezing and anti-squeezing ($r = 0$). The visibility at $r = 0$ has a value of $\sim 10^{-18}$ in Fig. 5, although that is not apparent from the plot. It can be seen that squeezing and anti-squeezing can produce a large improvement in the visibility under those conditions, although it cannot eliminate the decoherence altogether.

The effects of squeezing and anti-squeezing can be understood from Fig. 6, which is a plot of the mean number of photons left in the environment as a function of the squeezing parameter for the same conditions as in Fig. 5. The solid red line corresponds to the number of idler photons produced by the OPA, while the blue dotted line corresponds to the number of photons left in the environment by a beam splitter used to model the loss. It can be seen the optimal value of the squeezing parameter in Fig. 5 corresponds approximately to the minimum number of photons left in the environment in Fig. 6. It can also be seen that loss and gain leave approximately the same number of photons in the environment if $g = 1/t$. These results show once again that squeezing and anti-squeezing increase the visibility by reducing the number of photons left in the environment and thus the amount of which-path information.

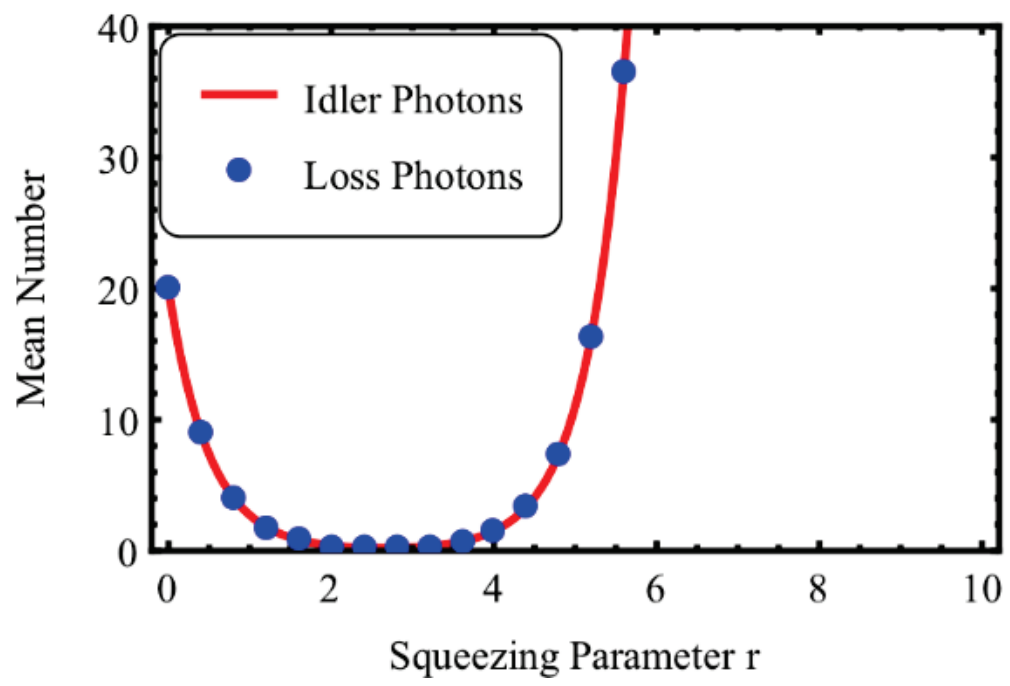

**Fig. 6.** (Color online) Plot of the average number of photons left in the environment as a function of the squeezing parameter $r$ under the same conditions used in Fig. 5. The solid red line corresponds to the number of idler photons emitted by the OPA, while the blue dots correspond to the number of photons left in the environment by a beam splitter used to model loss. The optimal value of the squeezing parameter r minimizes the number of photons left in the environment.

The visibility in Fig. 5 decreases after reaching a maximum value as the squeezing parameter is further increased. This can be understood from the fact that more squeezing will also increase the amplitude of the field along the imaginary axis in Fig. 2. The minimum transmission intensity and thus the maximum visibility occur when the real and imaginary components are approximately equal. That will be the case when the squeezing reduces the amplitude along the real axis from $\alpha_0$ to $\sqrt{\alpha_0}$, which increases the amplitude along the imaginary axis to $\sqrt{\alpha_0}$ from the uncertainty principle for the product of the two quadratures. Thus the optimal amount of squeezing reduces the amplitude by a factor $\sim\sqrt{\alpha_0}$ and it reduces the number of photons left in the environment by a factor $\sim|\alpha_0|$.

The decoherence due to entanglement with the idler photons in an OPA does not appear to be widely appreciated. It is a separate mechanism from the well-known quantum noise added by an amplifier. In fact, the decoherence from an OPA can be exponentially large even when the added quantum noise is negligible, as we showed in an earlier paper [5]. The quantum noise from the amplifier is included in these calculations, but its contribution to the decoherence is negligible compared to the which-path information in the limit of large $|\alpha_0|$ and small gain.

The visibility of the interference pattern depends on four parameters $\alpha_0$, $t$, $g$, and $r$, which results in a number of possible trade-offs in the choice of these parameters. If we assume that the gain is chosen to be $g = 1/t$ to restore the original amplitude of the quantum state, then the visibility only depends on three independent parameters. Here we will concentrate on the effects of gain, which have not been analyzed previously [20-21].

The maximum achievable visibility is plotted as a function of the initial amplitude $|\alpha_0|$ in Fig. 7 for several values of the gain ($g$ =1.001, 1.01, or 1.1). Here the optimal value of the squeezing parameter $r$ was calculated and used to evaluate the maximum visibility. Fig. 7a shows the optimal value of the squeezing parameter while Fig. 7b shows the corresponding visibility. It can be seen that the visibility decreases much faster as a function of $|\alpha_0|$ for large gains than it does for smaller gains. This is due to the fact that the average number of idler photons generated by an OPA is equal to $|\alpha_A|^2(g^2-1)$, where $\alpha_A$ is the amplitude of the field at the input to the amplifier [5]. Roughly speaking, the visibility will be substantially reduced when one or more idler photons are emitted on average, which corresponds to $|\alpha_A|^2(g^2-1)\sim 1$.

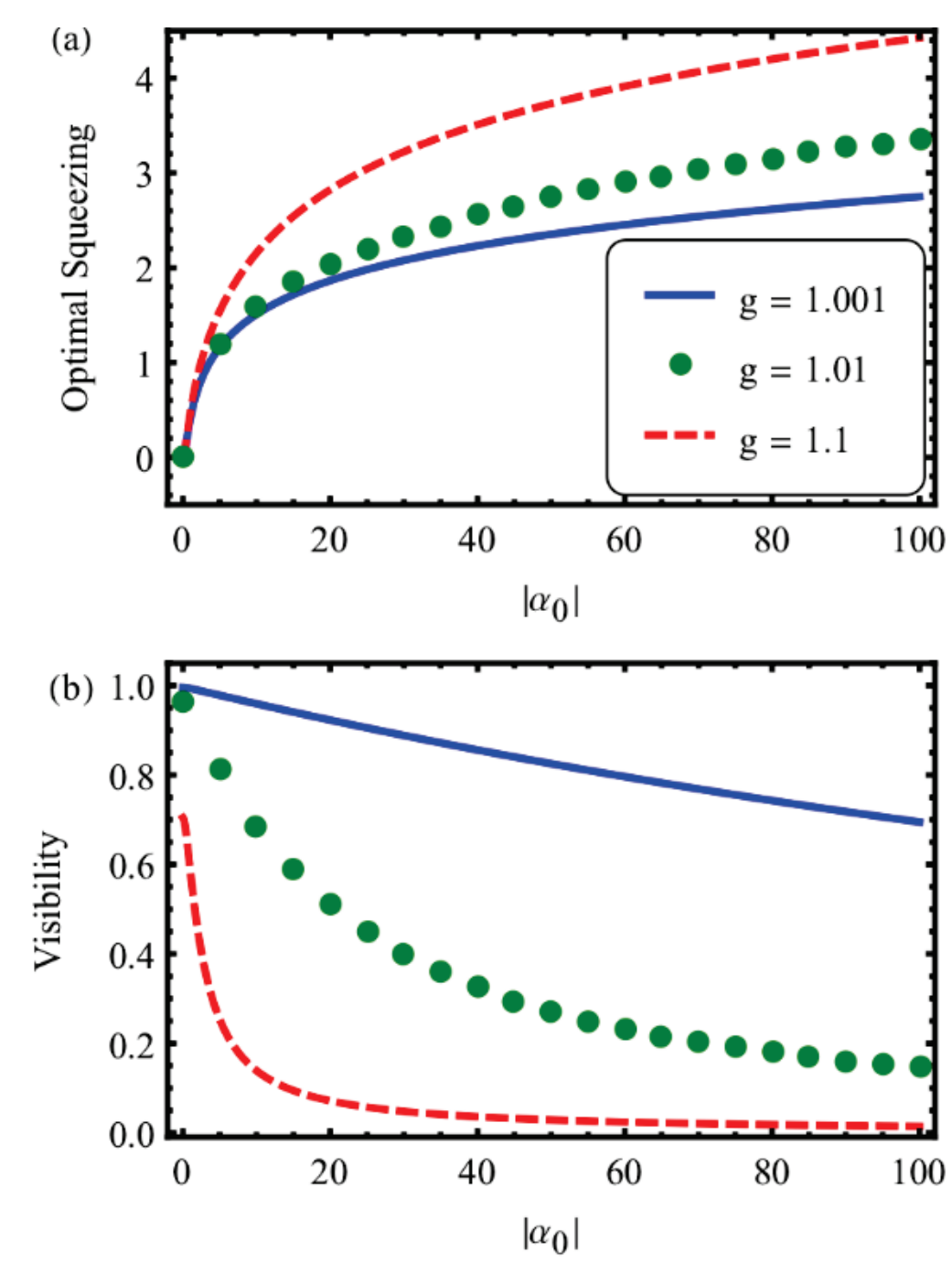

**Fig 7.** (Color online) Maximum achievable visibility as a function of the initial coherent state amplitude $|\alpha_0|$ for several values of the gain $g$. (a) The optimal value of the squeezing parameter $r$ as a function of $|\alpha_0|$. (b) The corresponding maximum visibility as a function of $|\alpha_0|$. It can be seen that the maximum visibility decreases more rapidly as a function of $|\alpha_0|$ for larger values of the gain.

The optimal squeezing parameter and the corresponding maximum visibility are plotted in Fig. 8 as a function of the gain for several values of $|\alpha_0|$ (1, 10, and 100). It can be seen once again that larger values of the gain require smaller values of $|\alpha_0|$ in order to achieve useful visibilities. The trade-off between the gain and $|\alpha_0|$ is further illustrated in Fig. 9, which shows the maximum visibility as a function of the gain or $|\alpha_0|$ for several values of the squeezing parameter $r$. It can be seen from Fig. 9b that a significant amount of visibility can be maintained even for relatively large mean photon numbers by using a sufficiently large amount of squeezing.

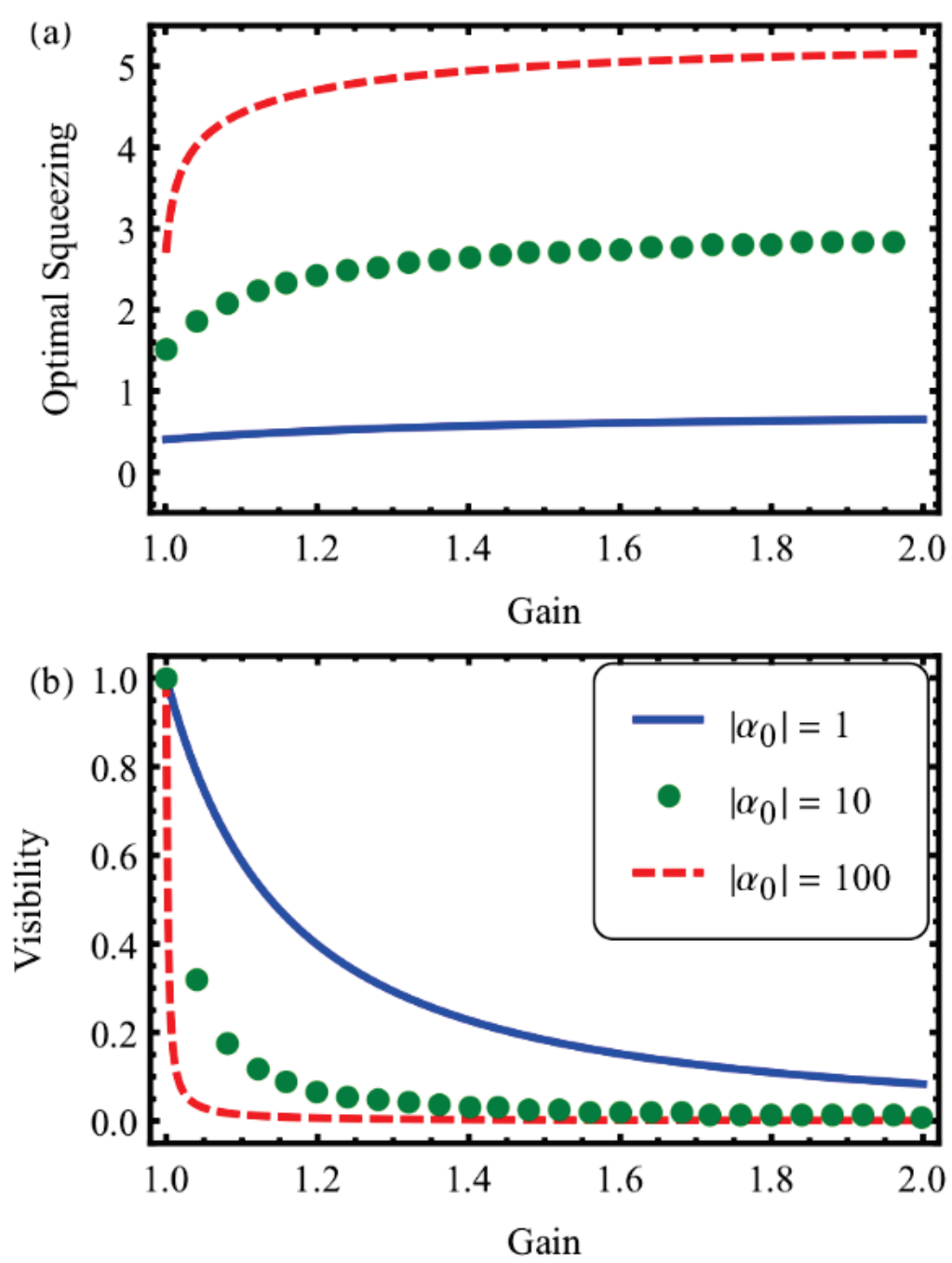

**Fig. 8.** (Color online) Optimal squeezing parameter $r$ and the corresponding maximum visibility as a function of the gain for several values of $|\alpha_0|$.

These results are summarized in the contour plots of Figs. 10 and 11. Fig. 10 shows the visibility as a function of $r$ and the gain for two values of the amplitude $|\alpha_0|$ (10 and 100), while Fig 11 shows the visibility as a function of $r$ and $|\alpha_0|$ for two values of the gain (1.01 and 1.1). It can be seen that relatively large visibilities can be obtained using squeezing and anti-squeezing as long as the product $|\alpha_0|(g^2-1)$ is less than unity, which corresponds to less than one idler photon on average. The relevance of this parameter can be seen in Fig. (10), where $|\alpha_0|$ is increased by a factor of 10 while $(g^2-1)$ is decreased by approximately a factor of 10 in going between Figs. (10a) and (10b), so that the product $|\alpha_0|(g^2-1)$ is essentially the same in the two parts of the figure. As a result, the visibility contours have roughly the same magnitude in Fig. (10b) as in Fig. (10a), although shifted to larger amounts of squeezing. Without any squeezing, the visibility would be seriously degraded when $|\alpha_0|^2(g^2-1)$ is on the order of unity instead. Thus squeezing and anti-squeezing can substantially reduce the amount of decoherence, but this approach is still limited to relatively small gains or initial amplitudes.

## V. SUMMARY AND CONCLUSIONS

In summary, we have shown that squeezing, amplification, and anti-squeezing can be used to reduce the decoherence of Schrodinger cat states during their transmission through a lossy medium, such as an optical fiber. Earlier studies [20-21] did not include the effects of amplification, which is required to restore the signal to its original amplitude.

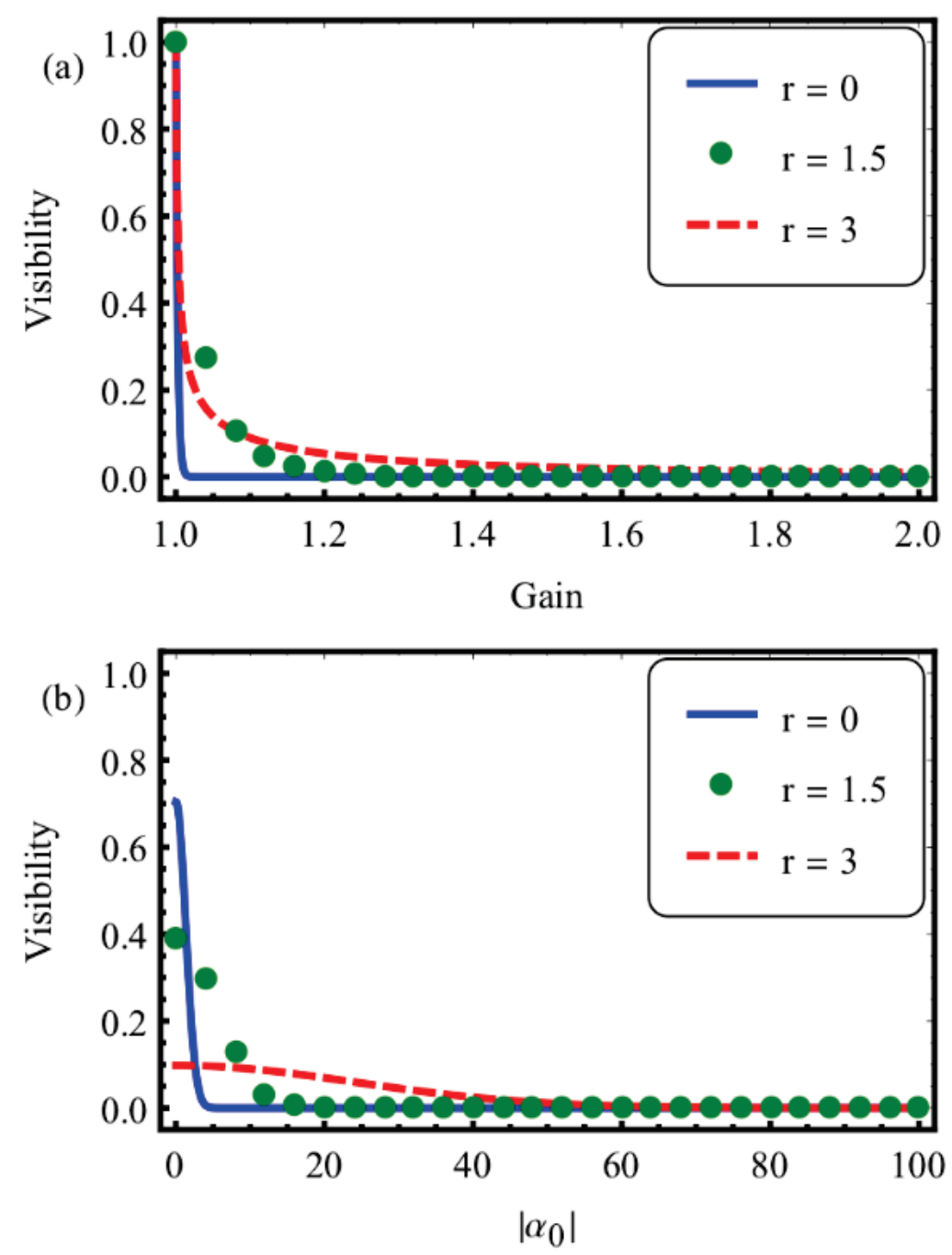

**Fig. 9.** (Color online) Plots of the visibility for several different values of the squeezing parameter $r$. (a) The visibility plotted as a function of the gain with $|\alpha_0|$ fixed at a value of 10. (b) The visibility as a function of $|\alpha_0|$ with the gain fixed at a value of 1.1.

The visibility of quantum interference effects without any squeezing and anti-squeezing can be exponentially small due to which-path information left in the environment by loss or amplification. Squeezing the state before transmission through a lossy channel can reduce the overall intensity of the signal and thus the decoherence due to which-path information left in the environment.

This approach has the advantage that it uses deterministic devices and does not reduce the data transmission rate as a result. On the other hand, the decoherence is not completely eliminated for any value of r, whereas the probabilistic approach of Ref. [23] can reduce the decoherence to an arbitrarily small amount at the expense of an exponentially small data transmission rate. In addition, this approach is only useful for non-Gaussian states with an asymmetrical Wigner distribution, such as a Schrodinger cat, whereas the approach of Ref. [23] can be applied to any state. Thus, there are several trade-offs to be considered in the use of these two approaches.

Macroscopic states and their decoherence mechanisms are a topic of fundamental interest, and our results provide further insight into this important topic. These results may also have practical applications in quantum sensor systems, for example, where the use of macroscopic superposition states may be beneficial in the presence of noise, and very high fidelities may not be required as is the case for quantum computing applications.

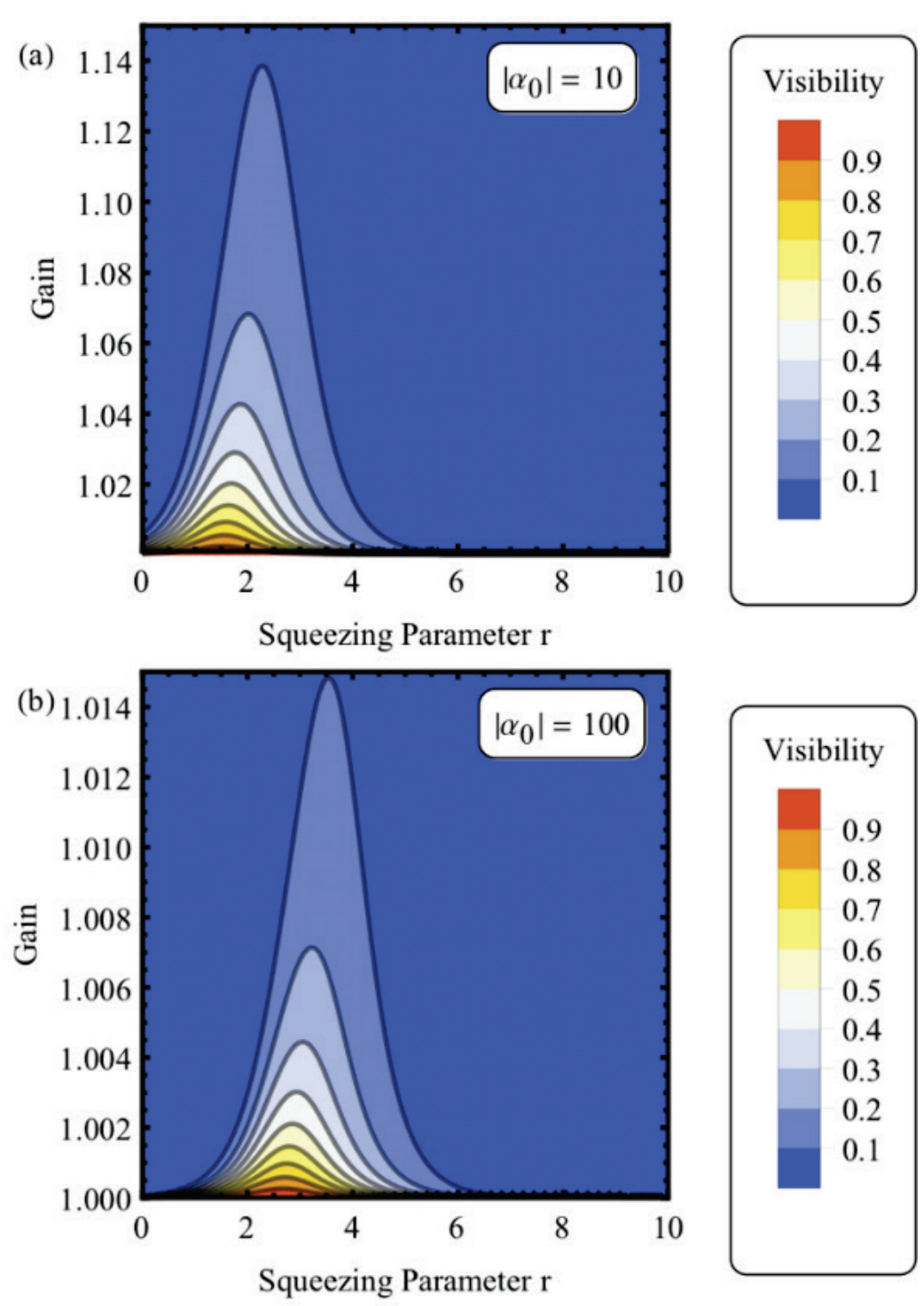

**Fig. 10.** (Color online) Contour plots of the visibility as a function of the amplitude gain and the squeezing parameter $r$. (a) $|\alpha_0| = 10$. (b) $|\alpha_0| = 100$.

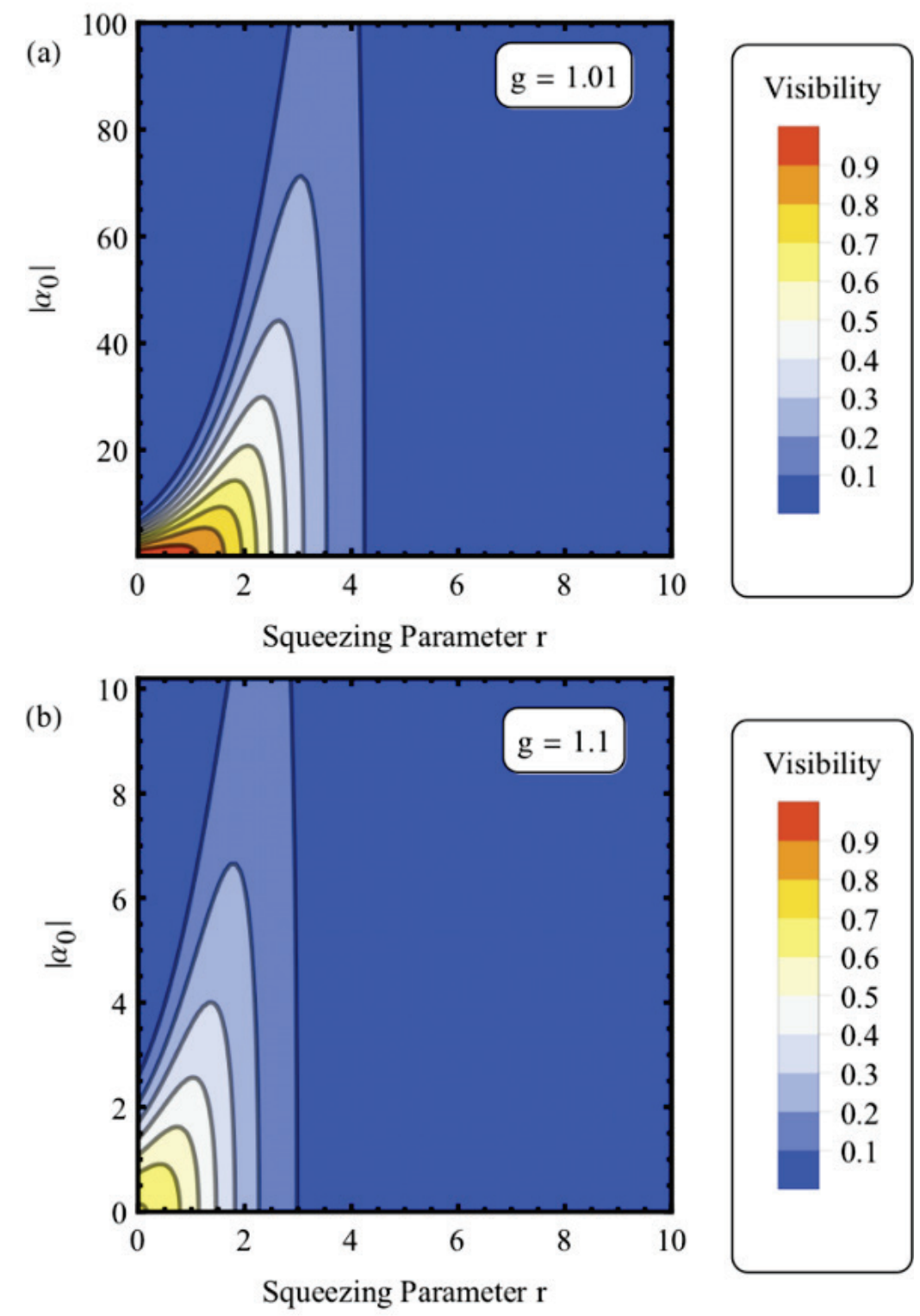

**Fig. 11.** (Color online) Contour plots of the visibility as a function of $|\alpha_0|$ and the squeezing parameter $r$. (a) Gain $g = 1.01$. (b) Gain $g = 1.1$.

**Acknowledgements**: This work was supported in part by a GAANN Fellowship from the U.S. Department of Education (P200A150003) and by the National Science Foundation under grant No. 1402708.

## References

[1] B. C. Sanders, Phys. Rev. A **45**, 6811 (1992).
[2] C. C. Gerry, Phys. Rev. A **59**, 4095 (1999).
[3] B. T. Kirby and J. D. Franson, Phys. Rev. A **87**, 053822 (2013).
[4] B. T. Kirby and J. D. Franson, Phys. Rev. A **89**, 033861 (2014).
[5] J. D. Franson and R. A. Brewster, Phys. Lett. A **382**, 887 (2018).
[6] R. J. Glauber, Phys. Rev. **131**, 2766 (1963).
[7] L. Mandel, Physica Scripta, **T12**, 34 (1985).
[8] R. E. Slusher, L.W. Hollberg, B. Yurke, J. C. Mertz and J. F. Valley, Phys. Rev. Lett. **55**, 2409 (1985).
[9] M.I. Kolobov, Rev. Mod. Phys. **71**, 1539 (1999).
[10] S. L. Braunstein and P. van Loock, Rev. Mod. Phys. **77**, 513 (2005).
[11] G. J. Milburn, M. L. Steyn-Ross and D. F. Walls, Phys. Rev. A **35**, 4443 (1987).
[12] Z. Y. Ou, S. F. Pereira and H. J. Kimble, Phys. Rev. Lett. **70**, 3239 (1993).
[13] C. M. Caves, Phys. Rev. Lett. **45**, 75 (1980).
[14] C. M. Caves, Phys. Rev. D **23**, 1693 (1981).
[15] R. S. Bondurant and J. H. Shapiro, Phys. Rev. D **30**, 2548 (1984).
[16] T. Eberle, S. Steinlechner, J. Bauchrowitz, V. Handchen. H. Vahlbruch, M. Mehmet, H. Muller-Ebhardt and Roman Schnabel, Phys. Rev. Lett. **104**, 251102 (2010).
[17] H. Vahlbruch, M. Mehmet, K. Danzmann and Roman Schnabel, Phys. Rev. Lett. **117**, 110801 (2016).
[18] C. C. Gerry and P. L. Knight, *Introductory Quantum Optics* (Cambridge University Press, Cambridge, 2005)**.**
[19] M. O. Scully and M. S. Zubairy, *Quantum Optics* (Cambridge University Press, New York, 1997).
[20] R. Filip, Phys. Rev. A **87**, 042308 (2013).
[21] H. Le Jeannic, A. Cavailles, K. Huang, R. Filip and J. Laurat, Phys. Rev. Lett. **120**, 073603 (2018); arXiv:1707.06244
[22] M. S. Kim and B. C. Sanders, Phys. Rev. A **53**, 3694-3697 (1996).
[23] M. Micuda, I. Straka, M. Mikova, M. Dusek, N. J. Cerf, J. Fiurasek and M. Jezek, Phys. Rev. Lett. **109**, 180503 (2012).
[24] C. N. Gagatsos, J. Fiurasek, A. Zavatta, M. Bellini and N. J. Cerf, Phys. Rev. A **89**, 062311 (2014).
[25] X. Zhao and G. Chiribella, Phys. Rev. A **35**, 8847 (2002).
[26] R. A. Brewster, I. C. Nodurft, T. B. Pittman and J. D. Franson, Phys. Rev. A **96**, 042307 (2017).
[27] G. Y. Xiang, T. C. Ralph, A. P. Lund, N. Walk and G. J. Pryde, Nat. Photonics **4**, 316 (2010).
[28] F. Ferreyrol, M. Barbieri, R. Blandino, S. Fossier, R. Tualle-Brouri and P. Grangier, Phys. Rev. Lett. **104**, 123603 (2010).
[29] A. Zavatta, J. Fiurasek and M. Bellini, Nat. Photonics **5**, 52 (2011).
[30] B. R. Mollow and R. J. Glauber, Phys. Rev. **160**, 1076-1096 (1967).
[31] B. R. Mollow and R. J. Glauber, Phys. Rev. **160**, 1097-1108 (1967).
[32] C. M. Caves, J. C. Combes, Z. Jiang and S. Pandey, Phys. Rev. A **86**, 063802 (2012).
[33] J. Niset, J. Fiurasek and N. J. Cerf, Phys. Rev. Lett. **102**, 120501 (2009).
[34] K. Husimi, Proc. Phys. Math. Soc. Jpn. **22**, 264 (1940).
[35] Y. Kano, J. Math. Phys. **6**, 1913 (1965).
[36] W. P. Schleich, *Quantum Optics in Phase Space* (WILEY-VCH, Berlin 2001).
[37] S. M. Barnett and P. L. Radmore, *Methods in Theoretical Quantum Optics* (Oxford University Press, New York, 1997).
[38] B. L. Schumaker and C. M. Caves, Phys. Rev. A **31**, 3093 (1985).
[39] A more detailed description of the necessary integrals can be found online in R.A. Brewster, T.B. Pittman, and J.D. Franson, arXiv:1803.00587v1.

**Appendix A: Q-function and visibility**

In this appendix, we will calculate the effects of the squeezing, loss, amplification, and anti-squeezing operations outlined in Fig. 3 of the main text. The corresponding unitary operators will be used to calculate the overall Q-function and the visibility of the Schrodinger-cat interferometer.

From Fig. 3 we see that the output state that arrives at the Homodyne detector is given by

$$|\psi\rangle_{\text{out}} = \hat{T}_2 \hat{S}^\dagger \hat{U}_a \hat{U}_b \hat{S} \hat{T}_1 \,|\,\alpha_0\rangle_s \,|\,0\rangle_e \,|\,0\rangle_i, \qquad \text{(A.1)}$$

where mode $s$ represents the signal mode which corresponds to annihilation operator $\hat{a}$, $e$ represents the ancillary mode of the loss operation (the environment) which corresponds to annihilation operator $\hat{b}$, and $i$ represents the idler mode of the parametric amplifier which corresponds to annihilation operator $\hat{c}$. The form of the various operators is described below. This expression differs from Eq. (7) in the text because loss and amplification have both been included here.

The effect of the first single-photon interferometer acting on the input coherent state is given by [5]

$$\hat{T}_1 = \frac{1}{2}\left(e^{i\phi\hat{a}^\dagger\hat{a}} + e^{i\theta}e^{-i\phi\hat{a}^\dagger\hat{a}}\right), \qquad \text{(A.2)}$$

which describes the creation of the Schrodinger cat state.

The squeezing operator $\hat{S}$ is given by [18, 19, 37]

$$\hat{S} = \frac{1}{\sqrt{\mu}} e^{-\frac{\sqrt{\mu^2-1}}{2\mu}e^{i\xi}\hat{a}^{\dagger 2}} \mu^{-\hat{a}^\dagger\hat{a}} e^{\frac{\sqrt{\mu^2-1}}{2\mu}e^{-i\xi}\hat{a}^2}, \qquad \text{(A.3)}$$

where $\mu$ is the hyperbolic cosine of the squeezing parameter $r$. This is equivalent to Eq. (4) in the text but it has been included here as well for completeness. The operator in Eq. (A.3) describes the initial single-mode squeezing while its inverse describes the anti-squeezing operation.

Loss can be modeled by a beam-splitter as is commonly done. The effects of the beam splitter correspond to a unitary operator $\hat{U}_b$ given by [37]

$$\hat{U}_b = e^{i\delta(\hat{a}^\dagger\hat{b} + \hat{a}\hat{b}^\dagger)}, \qquad \text{(A.4)}$$

where the transmission amplitude $t$ is given by $t = \cos\delta$ [37].

The effects of the parametric amplifier are described by the unitary operator $\hat{U}_a$ given by [32, 37-38]

$$\hat{U}_a = \frac{1}{g} e^{-\sqrt{g^2-1}\hat{a}^\dagger\hat{c}^\dagger/g} g^{-(\hat{a}^\dagger\hat{a}+\hat{c}^\dagger\hat{c})} e^{\sqrt{g^2-1}\hat{a}\hat{c}/g}, \qquad \text{(A.5)}$$

where g is the gain of the amplifier.

Finally, the second single photon interferometer is described by the operator [5]

$$\hat{T}_2 = \frac{1}{2}\left(e^{i\phi\hat{a}^\dagger\hat{a}} + e^{-i\phi\hat{a}^\dagger\hat{a}}\right), \qquad \text{(A.6)}$$

which has the same form as $\hat{T}_1$.

We will now calculate the Q-function corresponding to the final state of Eq. (A.1). The three mode Q-function can be defined as [34-36]

$$Q(\alpha,\beta,\gamma) = \frac{1}{\pi^3}\langle\gamma\,|_i\,\langle\beta\,|_e\,\langle\alpha\,|_s\;\hat{\rho}\,|\,\alpha\rangle_s\,|\,\beta\rangle_e\,|\,\gamma\rangle_i. \qquad \text{(A.7)}$$

Here $\alpha$ and $\gamma$ are the complex amplitudes of arbitrary coherent states in the signal and idler, respectively, while $\beta$ is the amplitude of an arbitrary coherent state in the other output of the beam splitter. From equation (A.1) the final density operator is given by

$$\begin{aligned}\hat{\rho}_{\text{out}} &= \hat{T}_2\hat{S}^\dagger\hat{U}_a\hat{U}_b\hat{S}\hat{T}_1\,|\,\alpha_0\rangle_s\,|\,0\rangle_e\,|\,0\rangle_i \\ &\times\langle 0\,|_i\,\langle 0\,|_e\,\langle\alpha_0\,|_s\;\hat{T}_1^\dagger\hat{S}^\dagger\hat{U}_b^\dagger\hat{U}_a^\dagger\hat{S}\hat{T}_2^\dagger.\end{aligned} \qquad \text{(A.8)}$$

Since we post-select on homodyne detector outputs that correspond to a net phase shift of zero as described in the main text, it will be convenient to define the variable $f_\sigma$ by

$$\begin{aligned} f_\sigma &\equiv \frac{\zeta_\sigma}{4}\langle\gamma\,|_i\,\langle\beta\,|_e\,\langle\alpha\,|_s\; e^{-i\sigma\hat{a}^\dagger\hat{a}}\hat{S}^\dagger\hat{U}_a\hat{U}_b\hat{S}e^{i\sigma\hat{a}^\dagger\hat{a}}\,|\alpha_0\rangle_s\,|0\rangle_e\,|0\rangle_i \\ &= \frac{\zeta_\sigma}{4}\langle\gamma\,|_i\,\langle\beta\,|_e\,\langle\alpha e^{i\sigma}\,|_s\;\hat{S}^\dagger\hat{U}_a\hat{U}_b\hat{S}\,|\,\alpha_0 e^{i\sigma}\rangle_s\,|0\rangle_e\,|0\rangle_i. \end{aligned} \qquad \text{(A.9)}$$

Here $\sigma$ is to be replaced with either $\phi$ or $-\phi$ as appropriate and

$$\zeta_\sigma \equiv \begin{cases} e^{i\theta} & \text{for} \quad \sigma = \phi \\ 1 & \text{for} \quad \sigma = -\phi \end{cases}. \qquad \text{(A.10)}$$

We can then define a general term in the post-selected Q-function as

$$Q_{\sigma,\tau}(\alpha,\beta,\gamma)=\frac{1}{\pi^3}f_\tau^* f_\sigma, \qquad \text{(A.11)}$$

which allows the full post-selected Q-function to be written as

$$\begin{aligned} Q(\alpha,\beta,\gamma) &= Q_{\phi,\phi}(\alpha,\beta,\gamma)+Q_{-\phi,-\phi}(\alpha,\beta,\gamma) \\ &+Q_{-\phi,\phi}(\alpha,\beta,\gamma)+Q_{\phi,-\phi}(\alpha,\beta,\gamma). \end{aligned} \qquad \text{(A.12)}$$

The value of $f_\sigma$ can be calculated using the completeness property of the coherent states:

$$\frac{1}{\pi}\int |\alpha\rangle\langle\alpha|\, d^2\alpha = \hat{1}. \qquad \text{(A.13)}$$

This allows Eq. (A.9) to be rewritten as

$$\begin{aligned} f_\sigma = \frac{\zeta_\sigma}{4\pi^9} &\int \langle\gamma|_i \langle\beta|_e \langle\alpha e^{i\sigma}|_s \hat{S}^\dagger |\varepsilon_1\rangle_s |\varepsilon_2\rangle_e |\varepsilon_3\rangle_i \\ &\times \langle\varepsilon_3|_i \langle\varepsilon_2|_e \langle\varepsilon_1|_s \hat{U}_a |\varepsilon_4\rangle_s |\varepsilon_5\rangle_e |\varepsilon_6\rangle_i \\ &\times \langle\varepsilon_6|_i \langle\varepsilon_5|_e \langle\varepsilon_4|_s \hat{U}_b |\varepsilon_7\rangle_s |\varepsilon_8\rangle_e |\varepsilon_9\rangle_i \\ &\times \langle\varepsilon_9|_i \langle\varepsilon_8|_e \langle\varepsilon_7|_s \hat{S} |\alpha_0 e^{i\sigma}\rangle_s |0\rangle_e |0\rangle_i \\ &\times d^2\varepsilon_1 d^2\varepsilon_2 d^2\varepsilon_3 d^2\varepsilon_4 d^2\varepsilon_5 d^2\varepsilon_6 d^2\varepsilon_7 d^2\varepsilon_8 d^2\varepsilon_9. \end{aligned} \qquad \text{(A.14)}$$

The factor involving $\hat{S}^\dagger$ in Eq. (A.14) can be reduced to

$$\begin{aligned} &\langle\gamma|_i \langle\beta|_e \langle\alpha e^{i\sigma}|_s \hat{S}^\dagger |\varepsilon_1\rangle_s |\varepsilon_2\rangle_e |\varepsilon_3\rangle_i \\ &= \frac{1}{\sqrt{\mu}} e^{-(|\alpha|^2+|\beta|^2+|\gamma|^2+|\varepsilon_1|^2+|\varepsilon_2|^2+|\varepsilon_3|^2)/2} \\ &\times e^{e^{-i\sigma}\alpha^*\varepsilon_1/\mu+\beta^*\varepsilon_2+\gamma^*\varepsilon_3} e^{\sqrt{\mu^2-1}e^{i(\xi-2\sigma)}\alpha^{*2}/2\mu-\sqrt{\mu^2-1}e^{-i\xi}\varepsilon_1^2/2\mu}. \end{aligned} \qquad \text{(A.15)}$$

Here we have used the adjoint of Eq. (A.3) and the fact that

$$\langle\alpha|\beta\rangle = e^{-(|\alpha|^2+|\beta|^2-2\alpha^*\beta)/2}. \qquad \text{(A.16)}$$

Next, Eqs. (A.5) and (A.16) give

$$\begin{aligned} &\langle\varepsilon_3|_i \langle\varepsilon_2|_e \langle\varepsilon_1|_s \hat{U}_a |\varepsilon_4\rangle_s |\varepsilon_5\rangle_e |\varepsilon_6\rangle_i \\ &= \frac{1}{g} e^{-(|\varepsilon_1|^2+|\varepsilon_2|^2+|\varepsilon_3|^2+|\varepsilon_4|^2+|\varepsilon_5|^2+|\varepsilon_6|^2)/2} \\ &\times e^{\varepsilon_1^*\varepsilon_4/g+\varepsilon_2^*\varepsilon_5+\varepsilon_3^*\varepsilon_6/g} e^{-\sqrt{g^2-1}\varepsilon_1^*\varepsilon_3^*/g+\sqrt{g^2-1}\varepsilon_4\varepsilon_6/g}. \end{aligned} \qquad \text{(A.17)}$$

The factor involving the operator $\hat{U}_b$ in Eq. (A.14) can be factored into a form similar to Eqs. (A.3) and (A.5) [37], but it is easier to use the fact that $\hat{U}_b|0\rangle = 0$ and [18, 19, 37]

$$\begin{aligned} \hat{U}_b \hat{a}^\dagger \hat{U}_b^\dagger &= t\hat{a}^\dagger + i\sqrt{1-t^2}\,\hat{b}^\dagger \\ \hat{U}_b \hat{b}^\dagger \hat{U}_b^\dagger &= t\hat{b}^\dagger + i\sqrt{1-t^2}\,\hat{a}^\dagger. \end{aligned} \qquad \text{(A.18)}$$

Using Eqs. (A.16) and (A.18) we have

$$\begin{aligned} &\langle\varepsilon_6|_i \langle\varepsilon_5|_e \langle\varepsilon_4|_s \hat{U}_a |\varepsilon_7\rangle_s |\varepsilon_8\rangle_e |\varepsilon_9\rangle_i \\ &= e^{-(|\varepsilon_4|^2+|\varepsilon_5|^2+|\varepsilon_6|^2+|\varepsilon_7|^2+|\varepsilon_8|^2+|\varepsilon_9|^2)/2} \\ &\times e^{t\varepsilon_4^*\varepsilon_7+t\varepsilon_5^*\varepsilon_8+\varepsilon_6^*\varepsilon_9} e^{i\sqrt{1-t^2}\varepsilon_4^*\varepsilon_8+i\sqrt{1-t^2}\varepsilon_5^*\varepsilon_7}. \end{aligned} \qquad \text{(A.19)}$$

The final factor in Eq. (A.14) can be evaluated by using Eqs. (A.3) and (A.16) again, which gives

$$\begin{aligned} &\langle\varepsilon_9|_i \langle\varepsilon_8|_e \langle\varepsilon_7|_s \hat{S} |\alpha_0 e^{i\sigma}\rangle_s |0\rangle_e |0\rangle_i \\ &= \frac{1}{\sqrt{\mu}} e^{-(|\varepsilon_7|^2+|\varepsilon_8|^2+|\varepsilon_9|^2+|\alpha_0|^2)/2} e^{e^{i\sigma}\varepsilon_7^*\alpha_0/\mu} \\ &\times e^{-\sqrt{\mu^2-1}e^{i\xi}\varepsilon_7^{*2}/2\mu+\sqrt{\mu^2-1}e^{-i(\xi-2\sigma)}\alpha_0^2/2\mu}. \end{aligned} \qquad \text{(A.20)}$$

Combining Eqs. (A.14), (A.15), (A.17), (A.19) and (A.20) gives the factor $f_\sigma$ in the form

$$\begin{aligned} f_\sigma &= \frac{\zeta_\sigma}{4\mu g\pi^9} e^{-(|\alpha|^2+|\beta|^2+|\gamma|^2+|\alpha_0|^2)/2} \\ &\times e^{\sqrt{\mu^2-1}e^{i(\xi-2\sigma)}\alpha^{*2}/2\mu+\sqrt{\mu^2-1}e^{-i(\xi-2\sigma)}\alpha_0^2/2\mu} \\ &\int e^{-(|\varepsilon_1|^2+|\varepsilon_2|^2+|\varepsilon_3|^2+|\varepsilon_4|^2+|\varepsilon_5|^2+|\varepsilon_6|^2+|\varepsilon_7|^2+|\varepsilon_8|^2+|\varepsilon_9|^2)} \\ &\times e^{e^{-i\sigma}\alpha^*\varepsilon_1/\mu+\beta^*\varepsilon_2+\gamma^*\varepsilon_3+\varepsilon_1^*\varepsilon_4/g+\varepsilon_2^*\varepsilon_5+\varepsilon_3^*\varepsilon_6/g+t\varepsilon_4^*\varepsilon_7+t\varepsilon_5^*\varepsilon_8+\varepsilon_6^*\varepsilon_9} \\ &\times e^{e^{i\sigma}\varepsilon_7^*\alpha_0/\mu-\sqrt{\mu^2-1}e^{-i\xi}\varepsilon_1^2/2\mu-\sqrt{g^2-1}\varepsilon_1^*\varepsilon_3^*/g+\sqrt{g^2-1}\varepsilon_4\varepsilon_6/g} \\ &\times e^{i\sqrt{1-t^2}\varepsilon_4^*\varepsilon_8+i\sqrt{1-t^2}\varepsilon_5^*\varepsilon_7-\sqrt{\mu^2-1}e^{i\xi}\varepsilon_7^{*2}/2\mu} \\ &\times d^2\varepsilon_1 d^2\varepsilon_2 d^2\varepsilon_3 d^2\varepsilon_4 d^2\varepsilon_5 d^2\varepsilon_6 d^2\varepsilon_7 d^2\varepsilon_8 d^2\varepsilon_9. \end{aligned} \qquad \text{(A.21)}$$

The higher-dimensional Gaussian integral in Eq. (A.21) looks complicated but it can be evaluated using an appropriate change of variables. As described in more detail in an earlier online version of this paper [39], the result is that

$$\begin{aligned}
f_\sigma = {} & \frac{\zeta_\sigma e^{-(|\alpha|^2+|\beta|^2+|\gamma|^2+|\alpha_0|^2)/2}}{4\sqrt{\mu^2 g^2 - t^2(\mu^2-1)}} \\
& \times e^{\mu\sqrt{\mu^2-1}(g^2-t^2)e^{-i(\xi-2\sigma)}\alpha_0^2/2[\mu^2g^2-t^2(\mu^2-1)]} \\
& \times e^{\mu\sqrt{\mu^2-1}(g^2-t^2)e^{i(\xi-2\sigma)}\alpha^{*2}/2[\mu^2g^2-t^2(\mu^2-1)]} \\
& \times e^{(1-t^2)g^2\mu\sqrt{\mu^2-1}e^{i\xi}\beta^{*2}/2[\mu^2g^2-t^2(\mu^2-1)]} \\
& \times e^{\mu\sqrt{\mu^2-1}(g^2-1)e^{-i\xi}\gamma^{*2}/2[\mu^2g^2-t^2(\mu^2-1)]} \\
& \times e^{tg\alpha^*\alpha_0/[\mu^2g^2-t^2(\mu^2-1)]+i\sqrt{1-t^2}g^2\mu e^{i\sigma}\beta^*\alpha_0/[\mu^2g^2-t^2(\mu^2-1)]} \\
& \times e^{t\sqrt{g^2-1}\sqrt{\mu^2-1}e^{-i(\xi-\sigma)}\gamma^*\alpha_0/[\mu^2g^2-t^2(\mu^2-1)]} \\
& \times e^{-it\sqrt{1-t^2}g\sqrt{\mu^2-1}e^{i(\xi-\sigma)}\alpha^*\beta^*/[\mu^2g^2-t^2(\mu^2-1)]} \\
& \times e^{-\mu g\sqrt{g^2-1}e^{-i\sigma}\alpha^*\gamma^*/[\mu^2g^2-t^2(\mu^2-1)]} \\
& \times e^{-it\sqrt{1-t^2}\sqrt{g^2-1}(\mu^2-1)\beta^*\gamma^*/[\mu^2g^2-t^2(\mu^2-1)]},
\end{aligned} \tag{A.22}$$

Inserting Eq. (A.22) into (A.11) gives the general term of the Q-function as

$$\begin{aligned}
Q_{\sigma,\tau}(\alpha,\beta,\gamma) = {} & \frac{\zeta_\tau^*\zeta_\sigma e^{-(|\alpha|^2+|\beta|^2+|\gamma|^2+|\alpha_0|^2)}}{16\pi^3[\mu^2 g^2 - t^2(\mu^2-1)]} \\
& \times e^{\mu\sqrt{\mu^2-1}(g^2-t^2)[e^{-i(\xi-2\sigma)}\alpha_0^2+e^{i(\xi-2\tau)}\alpha_0^{*2}]/2[\mu^2g^2-t^2(\mu^2-1)]} \\
& \times e^{\mu\sqrt{\mu^2-1}(g^2-t^2)[e^{-i(\xi-2\tau)}\alpha^2+e^{i(\xi-2\sigma)}\alpha^{*2}]/2[\mu^2g^2-t^2(\mu^2-1)]} \\
& \times e^{(1-t^2)g^2\mu\sqrt{\mu^2-1}[e^{-i\xi}\beta^2+e^{i\xi}\beta^{*2}]/2[\mu^2g^2-t^2(\mu^2-1)]} \\
& \times e^{\mu\sqrt{\mu^2-1}(g^2-1)[e^{i\xi}\gamma^2+e^{-i\xi}\gamma^{*2}]/2[\mu^2g^2-t^2(\mu^2-1)]} \\
& \times e^{tg[\alpha^*\alpha_0+\alpha_0^*\alpha]/[\mu^2g^2-t^2(\mu^2-1)]} \\
& \times e^{i\sqrt{1-t^2}g^2\mu[e^{i\sigma}\beta^*\alpha_0+e^{-i\sigma}\alpha_0^*\beta]/[\mu^2g^2-t^2(\mu^2-1)]} \\
& \times e^{t\sqrt{g^2-1}\sqrt{\mu^2-1}[e^{-i(\xi-\sigma)}\gamma^*\alpha_0+e^{i(\xi-\tau)}\alpha_0^*\gamma]/[\mu^2g^2-t^2(\mu^2-1)]} \\
& \times e^{it\sqrt{1-t^2}g\sqrt{\mu^2-1}[e^{-i(\xi-\tau)}\alpha\beta-e^{i(\xi-\sigma)}\alpha^*\beta^*]/[\mu^2g^2-t^2(\mu^2-1)]} \\
& \times e^{-\mu g\sqrt{g^2-1}[e^{i\tau}\alpha\gamma+e^{-i\sigma}\alpha^*\gamma^*]/[\mu^2g^2-t^2(\mu^2-1)]} \\
& \times e^{it\sqrt{1-t^2}\sqrt{g^2-1}(\mu^2-1)[\beta\gamma-\beta^*\gamma^*]/[\mu^2g^2-t^2(\mu^2-1)]},
\end{aligned} \tag{A.23}$$

with the full Q-function being given by Eq. (A.12).

We can now calculate the visibility $V$ of the quantum interference, which is defined as $V \equiv (P_{\max} - P_{\min})/(P_{\max} + P_{\min})$. Here $P_{\max}$ and $P_{\min}$ are the maximum and minimum probabilities obtained by varying the phase $\theta$ of the phase shifter in the first single-photon interferometer. In order to do this, we use the fact that the total probability $P$ of a post-selected event is given by [5, 34-36]

$$P = \int d^2\alpha\, d^2\beta d^2\gamma\, Q(\alpha,\beta,\gamma). \tag{A.24}$$

By inspection of equations (A.12) and (A.23) we see that the visibility is given by

$$\begin{aligned}
v &= \left| \int Q_{\phi,-\phi}(\alpha,\beta,\gamma) d^2\gamma d^2\beta d^2\alpha \right| \\
&= \left| \int Q_{-\phi,\phi}(\alpha,\beta,\gamma) d^2\gamma d^2\beta d^2\alpha \right|.
\end{aligned} \tag{A.25}$$

Unlike Eq. (A.21), the integrals in Eq. (A.24) are relatively complicated. The same change of variables can still be used if we write the complex parameters in terms of their real and imaginary parts:

$$\alpha = \alpha_r + i\alpha_i, \tag{A.26}$$

and

$$d^2\alpha = d\alpha_r d\alpha_i, \tag{A.27}$$

for the signal mode. Similar expressions exist for $\beta$ and $\gamma$ of the environment and idler modes, respectively. This allows the integral in Eq. (A.24) to be evaluated analytically, although the resulting equations are very lengthy and not included here. Examples of the resulting visibility are plotted in Section IV of the text.

**Appendix B: Idler photons created in the amplification of a squeezed coherent state**

As noted in the main text, the reduction in the decoherence from the squeezing and anti-squeezing operations is due to a decrease in the number of photons left in the environment, which reduces the amount of which-path information. We will illustrate this by calculating the number of photons produced during the amplification process, as shown in Fig. 6 of the main text.

Consider a general input state expanded in a basis of number states:

$$|\psi\rangle_{\text{in}} = \sum_{n=0}^{\infty} c_n |n\rangle_s |0\rangle_i. \tag{B.1}$$

Here the $c_n$ are the probability amplitudes for number state $|n\rangle$ and the idler is assumed to initially be in its vacuum state. Applying the evolution operator for a parametric amplifier $\hat{U}$, given by Eq. (5) of the main text or Eq. (A.5), to a number state can be shown to give

$$\hat{U}\,|\,n\rangle_s\,|\,0\rangle_i = \frac{1}{g^{n+1}}\sum_{j=0}^{\infty}\sqrt{\frac{(j+n)!}{j!}}\left(-\frac{\sqrt{g^2-1}}{g}\right)^j |\,j+n\rangle_s\,|\,j\rangle_i. \qquad \text{(B.2)}$$

Thus, we have

$$\hat{U}\,|\,\psi\rangle_{\text{in}} = \frac{1}{g}\sum_{n=0}^{\infty}\frac{c_n}{g^n\sqrt{n!}}\sum_{j=0}^{\infty}\sqrt{\frac{(j+n)!}{j!}}\left(-\frac{\sqrt{g^2-1}}{g}\right)^j |\,j+n\rangle_s\,|\,j\rangle_i. \qquad \text{(B.3)}$$

We can calculate the two-mode Q-function using [34-36]

$$Q(\alpha,\beta) = \frac{1}{\pi^2}\langle\beta\,|_i\,\langle\alpha\,|_s\,\hat{U}\,|\,\psi\rangle_{\text{in}} \times\langle\psi\,|_{\text{in}}\,\hat{U}^\dagger\,|\,\alpha\rangle_s\,|\,\beta\rangle_i. \qquad \text{(B.4)}$$

Combining Eqs. (B.3) and (B.4), using Eq. (A.16) and performing the sum over j we get the Q-function in the form

$$Q(\alpha,\beta) = \frac{1}{\pi^2 g^2}e^{-(|\alpha|^2+|\beta|^2)}e^{-\sqrt{g^2-1}(\alpha\beta+\alpha^*\beta^*)/g} \times\sum_{n=0}^{\infty}\sum_{m=0}^{\infty}\frac{c_m^* c_n}{g^{n+m}\sqrt{n!m!}}\alpha^{*n}\alpha^m. \qquad \text{(B.5)}$$

We now trace over the signal mode, which is equivalent to integrating over the real and imaginary parts of $\alpha$ in the Q-function. Performing the integral gives the reduced Q-function $Q(\beta)$ in the form

$$Q(\beta) = \frac{e^{-|\beta|^2/g}}{\pi g^2}\sum_{n=0}^{\infty}\sum_{m=0}^{\infty}\frac{c_m^* c_n}{g^{n+m}}\sqrt{\frac{n!}{m!}}\left(-\frac{\sqrt{g^2-1}}{g}\beta^*\right)^{m-n} \times L_n^{m-n}\left[-\frac{g^2-1}{g^2}|\,\beta\,|^2\right], \qquad \text{(B.6)}$$

where $L_n^a$ are the associated Laguerre polynomials.

Our goal is to calculate the number of photons left in the idler mode. The expectation value $\langle\hat{n}_i\rangle$ is given by

$$\langle\hat{n}_i\rangle = \langle\hat{b}^\dagger\hat{b}\rangle = \langle\hat{b}\hat{b}^\dagger - 1\rangle = \int Q(\beta)(|\,\beta\,|^2 - 1)d^2\beta, \qquad \text{(B.7)}$$

where the integral is to be performed over all values of the real and imaginary parts of $\beta$. Here we have made use of the fact that the Q-function is an antinormally ordered quasiprobability distribution [34-36].

Inserting (B.6) into (B.7) and evaluating the integral gives the simple expression

$$\langle\hat{n}_i\rangle = (g^2-1)\sum_{n=0}^{\infty}(1+n)\,|\,c_n\,|^2. \qquad \text{(B.8)}$$

Note that any terms with $m \neq n$ do not contribute to the integral. Equation (B.8) gives the average number of idler photons generated by amplification of the general state given in equation (B.1).

We are interested in the number of idler photons created by amplifying a squeezed coherent state given by

$$|\,\psi\rangle_{\text{in}} = \hat{S}\,|\,\alpha_0\rangle_s\,|\,0\rangle_i \qquad \text{(B.9)}$$

where $\hat{S}$ is the single-mode squeeze operator given by Eqs. (3) or (4) in the main text or Eq. (B.3). It can be shown that

$$\hat{S}\left|\alpha_0\right\rangle_s = \frac{e^{-\left|\mu\alpha_0-\sqrt{\mu^2-1}e^{i\xi}\alpha_0^*\right|^2/2}}{\sqrt{\mu}} \times\sum_{j=0}^{\infty}\frac{(2j)!}{j!}\left(-\frac{e^{i\xi}\sqrt{\mu^2-1}}{2\mu(\mu\alpha_0-\sqrt{\mu^2-1}e^{i\xi}\alpha_0^*)^2}\right)^j \times\sum_{k=0}^{\infty}\frac{(\mu\alpha_0-\sqrt{\mu^2-1}e^{i\xi}\alpha_0^*)^k}{\sqrt{k!}} \times L_{2j}^{k-2j}\left[\left|\mu\alpha_0-\sqrt{\mu^2-1}e^{i\xi}\alpha_0^*\right|^2\right]\left|k\right\rangle \qquad \text{(B.10)}$$

Inserting (B.10) into (B.9), using the fact that $c_n = \langle n\,|\,\psi\rangle_{\text{in}}$, and performing the sum over j and k gives

$$c_n = \frac{e^{-|\alpha_0|^2/2}e^{-\sqrt{\mu^2-1}e^{-i\xi}\alpha_0^2/2\mu}}{\sqrt{\mu n!}}\left(\frac{\sqrt{\mu^2-1}e^{i\xi}}{2\mu}\right)^{n/2} \times H_n\left[\frac{\alpha_0 e^{-i\xi/2}}{\sqrt{2\mu\sqrt{\mu^2-1}}}\right], \qquad \text{(B.11)}$$

where $H_n$ are the Hermite polynomials.

To find the number of idler photons generated by the amplification of a squeezed coherent state, we insert Eq. (B.11) into (B.8) and perform the sum over n. The result is that

$$\langle \hat{n}_i \rangle = (g^2 - 1)\Big((2\mu^2 - 1)\,|\,\alpha_0\,|^2 + \mu^2 \\ -\mu\sqrt{\mu^2 - 1}(e^{-i\xi}\alpha_0^2 + e^{i\xi}\alpha_0^{*2})\Big). \tag{B.12}$$

Eq. (B.12) is plotted in Fig. 6 in the text for several sets of parameters. It can be seen that squeezing an input coherent state with an appropriate value of $\xi$ and the squeezing parameter can reduce the number of idler photons created during amplification. The inclusion of loss as well as amplification gives a similar result.